\documentclass[b5paper,twoside,reqno]{bjp}
\usepackage{cite}
\usepackage{latexsym}
\usepackage{graphicx}
\usepackage{amssymb,amsmath,amsfonts,amsbsy, amsthm, xspace, latexsym, amscd, enumitem}
\usepackage[utf8]{inputenc}
\usepackage[T1]{fontenc}
\usepackage[english]{babel}
\usepackage{times}
\usepackage{url,bm}
\newcommand\rf[1]{(\ref{eq:#1})}
\newcommand\lab[1]{\label{eq:#1}}
\newcommand\nonu{\nonumber}
\newcommand\br{\begin{eqnarray}}
\newcommand\er{\end{eqnarray}}
\newcommand\be{\begin{equation}}
\newcommand\ee{\end{equation}}

\newcommand\foot[1]{\footnotemark\footnotetext{#1}}
\newcommand\lb{\lbrack}
\newcommand\rb{\rbrack}

\newcommand\llb{\left\lbrack}
\newcommand\rrb{\right\rbrack}

\renewcommand\({\left(}
\renewcommand\){\right)}
\newcommand\bv{\bigm\vert}               

\newcommand\bc{\begin{center}}
\newcommand\ec{\end{center}}

\newcommand\partder[2]{\frac{{\partial {#1}}}{{\partial {#2}}}}

\renewcommand\a{\alpha}

\renewcommand\d{\delta}

\newcommand\eps{\epsilon}
\newcommand\vareps{\varepsilon}

\newcommand\G{\Gamma}

\newcommand\h{\frac{1}{2}}
\renewcommand\k{\kappa}
\renewcommand\l{\lambda}
\renewcommand\L{\Lambda}
\newcommand\m{\mu}
\newcommand\n{\nu}

\newcommand\vp{\varphi}
\renewcommand\P{\Phi}
\newcommand\pa{\partial}

\newcommand\pr{\prime}

\newcommand\s{\sigma}

\renewcommand\t{\tau}

\newcommand\wti{\widetilde}

\newcommand\cA{{\mathcal A}}
\newcommand\cB{{\mathcal B}}
\newcommand\cC{{\mathcal C}}

\newcommand{\ct}[1]{\cite{#1}}
\newcommand{\bib}[1]{\bibitem{#1}}

\newcommand\vpdot{\stackrel{.}{\varphi}}
\newcommand\vpddot{\stackrel{..}{\varphi}}

\newcommand\adot{\stackrel{.}{a}}
\newcommand\addot{\stackrel{..}{a}}

\renewcommand\congress{\small\sf Section: Mathematical Physics}

\makeindex

\begin{document}

\sloppy \raggedbottom

\title{Quintessential Inflation, Unified Dark Energy and Dark Matter, and Higgs Mechanism}

\runningheads{Quintessence, Unified Dark Energy and Dark Matter, and Higgs Mechanism}{E.
Guendelman, E. Nissimov and S. Pacheva}

\begin{start}
\coauthor{Eduardo Guendelman}{1}, \author{Emil Nissimov}{2}, 
\author{Svetlana Pacheva}{2}

\address{Department of Physics, Ben-Gurion University of the Negev, \\
Beer-Sheva 84105, Israel}{1}

\address{Institute for Nuclear Research and Nuclear Energy,\\ 
Bulgarian Academy of Sciences, Sofia 1784, Bulgaria}{2}

\begin{Abstract}
We describe a new type of gravity-matter models where gravity couples in a
non-conventional way to two distinct scalar fields providing a unified 
Lagrangian action principle description of:
(a) the evolution of both ``early'' and ``late'' Universe -- by the
``inflaton'' scalar field;
(b) dark energy and dark matter as a unified manifestation of a single
material entity - the ``darkon'' scalar field.
The essential non-standard feature of our models is employing the formalism of 
non-Riemannian space-time volume forms -- alternative generally covariant 
integration measure densities (volume elements)  defined in terms of auxiliary
antisymmetric tensor gauge fields. 
Although being (almost) pure-gauge degrees of freedom, the non-Riemannian space-time 
volume forms trigger a series of important features unavailable in ordinary 
gravity-matter models 
When including in
addition interactions with the electro-weak model bosonic sector 
we obtain a  gravity-assisted generation of electro-weak spontaneous gauge symmetry
breaking in the post-inflationary ``late'' Universe, while the Higgs-like
scalar remains massless in the ``early'' Universe.
\end{Abstract}

\end{start}

\section{Introduction}
\label{intro}

Dark energy and dark matter, occupying around 70\% and 25\% of the matter content of 
the Universe, respectively, continue to be the two most unexplained ``mysteries''
in cosmology and astrophysics (for a background, see \ct{DE-rev,DM-rev}).
In most loose terms dark energy is responsible for 
the accelerated expansion of today's Universe, i.e., dark energy acts effectively 
as repulsion force among the galaxies -- a phenomenon completely counterintuitive 
w.r.t. the naive notion about gravity as an attractive force. And vice versa, dark matter 
holds together the matter objects inside the galaxies. The adjective ``dark'' is due 
to the fact that both these fundamental matter components of the Universe interact 
only gravitationally, and they do not directly interact with ordinary (baryonic) 
matter, in particular, they do not interact electromagnetically and thus they remain 
``dark''.

There exist a multitude of proposals for an adequate description of dark energy's 
and dark matter's dynamics within the framework of standard general relativity 
or its modern extensions \ct{chaplygin,purely-kinetic-k-essence,mimetic}.
Here we will briefly describe and further extend the basic features of our
approach to the above topic \ct{dusty-dusty-2} (for some earlier works, see
also \ct{eduardo-singleton-ansoldi}).
   
Using the method of non-Riemannian spacetime volume-forms (metric-independent 
generally-covariant integration measure densities or volume elements) \ct{TMT-orig}
we start by constructing from first principles (via Lagrangian action) a new 
non-canonical cosmological model of gravity interacting with a single scalar 
field (here called ``darkon''), which explicitly yields a self-consistent unified description 
of dark energy as a dynamically generated cosmological constant, and 
dark matter as a dust fluid flowing along spacetime geodesics, by unifying 
them as an exact sum of two separate contributions to the pertinent scalar field 
energy-momentum tensor. In other words, this unified description shows that
dark energy and dark matter may be viewed as two different manifestations of one 
single matter source - the scalar ``darkon'' field \ct{dusty-dusty-2}.
    
Next, extending our formalism of non-Riemannian spacetime volume-forms, we couple 
the above non-canonical gravity-matter system to a second scalar field -- 
the ``inflaton'' -- in such a way that the ``inflaton'' dynamics provides a unified
description of the evolution of both ``early'' and ``late'' Universe
\ct{emergent} -- this is a model of ``quintessential inflation'' \ct{quintessence-orig}. 
Furthermore, we add interaction with the $SU(2)\times U(1)$
scalar and gauge fields of the electro-weak bosonic sector.

We exhibit in some detail the interplay between the ``inflaton'' and the ``darkon'' in the
``early'' (inflationary)  and the ``late'' (dark energy dominated) epochs of the Universe.
Among the principal interesting features is the gravity-assisted generation in the ``late'' 
Universe of Higgs-like spontaneous gauge symmetry breaking effective potential for 
the $SU(2)\times U(1)$ scalar iso-doublet \foot{For a related approach, 
see \ct{grf-essay} based on an old idea by Bekenstein \ct{bekenstein}.}. In
the ``early'' Universe the Higgs-like field remains massless.

\section{Hidden Noether Symmetry and Unification of Dark Energy and Dark Matter}
\label{DE-DM}

First we will consider, following \ct{dusty-dusty-2}, a simple particular case of a 
non-conventional gravity-scalar-field action -- a member of the general class of the 
``modified-measure'' gravity-matter theories \ct{TMT-orig}
(for simplicity we use units with the Newton constant $G_N = 1/16\pi$):
\be
S = \int d^4 x \sqrt{-g}\, R +
\int d^4 x \bigl(\sqrt{-g}+\P(C)\bigr) L(u,Y) \; .
\lab{TMT-0}
\ee
Here $R$ denotes the standard Riemannian scalar curvature for the pertinent
Riemannian metric $g_{\m\n}$. 
In the second term in \rf{TMT-0} -- the scalar field Lagrangian is coupled
{\em symmetrically} to two mutually independent spacetime volume-forms 
(integration measure densities or volume elements) 
-- the standard Riemannian $\sqrt{-g} = \sqrt{-\det\Vert g_{\m\n}\Vert}$ and to an 
alternative non-Riemannian one:
\be
\P(C) = \frac{1}{3!}\vareps^{\m\n\k\l} \pa_\m C_{\n\k\l} \; ,
\lab{mod-measure}
\ee
where $C_{\m\n\l}$ is an auxiliary rank 3 antisymmetric tensor field.

$L(u,Y)$ is general-coordinate invariant Lagrangian of a single scalar field 
$u (x)$:
\be
L(u,Y) = Y - V(u) \quad ,\quad Y \equiv - \h g^{\m\n}\pa_\m u \pa_\n u \; .
\lab{standard-L}
\ee
As a result of the equations of motion w.r.t. ``measure'' gauge field $C_{\m\n\l}$
we obtain the following crucial new property of the model \rf{TMT-0}
-- {\em dynamical constraint} on $L(u,Y)$ alongside with the second-order
differential equations of motion w.r.t. $u$ (which now contains the
non-Riemannian volume element $\P(C)$ \rf{mod-measure}):
\be
\pa_\m L (u,Y) = 0 \;\;\; \longrightarrow \;\;\;
L (u,Y) = - 2M_0 = {\rm const} \;\; ,\;\; {\rm i.e.}\;\; Y = V(u) - 2M_0 \; ,
\lab{L-const}
\ee
where $M_0$ is arbitrary integration constant. The factor $2$ in front of $M_0$ is 
for later convenience in view of its interpretation as a 
{\em dynamically generated cosmological constant}. 

Indeed, taking into account \rf{L-const} the energy-momentum tensor becomes:
\be
T_{\m\n} = -2 g_{\m\n} M_0 + 
\Bigl( 1+\frac{\P(C)}{\sqrt{-g}}\Bigr) \pa_\m u\, \pa_\n u 
\quad ,\quad \nabla^\n T_{\m\n} = 0 \; .
\lab{EM-tensor}
\ee

A second crucial property of the model \rf{TMT-0} is the existence of a
{\em hidden strongly nonlinear Noether symmetry}
due to the presence of the non-Riemannian volume element $\P (C)$:
\br
\d_\eps u = \eps \sqrt{Y} \;\;\; ,\;\;\; \d_\eps g_{\m\n} = 0 \;\;\; ,\;\;\;
\d_\eps \cC^\m = - \eps \frac{1}{2\sqrt{Y}} g^{\m\n}\pa_\n u 
\bigl(\P(C) + \sqrt{-g}\bigr)  \; ,
\lab{hidden-sym}
\er
where $\cC^\m \equiv \frac{1}{3!} \vareps^{\m\n\k\l} C_{\n\k\l}$.
Under \rf{hidden-sym}  the action \rf{TMT-0} transforms as:
$\d_\eps S = \int d^4 x\, \pa_\m \bigl( L(u,Y) \d_\eps \cC^\m \bigr)$ .
Then, standard Noether procedure yields a conserved current:
\be
\nabla_\m J^\m = 0 \quad ,\quad
J^\m \equiv - \Bigl(1+\frac{\P(C)}{\sqrt{-g}}\Bigr)\sqrt{2Y} g^{\m\n}\pa_\n u \; .
\lab{J-conserv}
\ee

$T_{\m\n}$ \rf{EM-tensor} and $J^\m$ \rf{J-conserv} can be cast into
a relativistic hydrodynamical form (taking into account \rf{L-const}):
\be
T_{\m\n} = - 2M_0 g_{\m\n} + \rho_0 u_\m u_\n \quad ,\quad J^\m = \rho_0 u^\m \; ,
\lab{T-J-hydro}
\ee
where:
\be
\rho_0 \equiv \Bigl(1+\frac{\P(C)}{\sqrt{-g}}\Bigr)\, 2Y 
\;\;\; ,\;\; 
u_\m \equiv - \frac{\pa_\m u}{\sqrt{2Y}} \;\; ,\;\; 
 u^\m u_\m = -1 \; .
\lab{rho-0-def}
\ee
For the pressure $p$ and energy density $\rho$ we have accordingly
(with $\rho_0$ as in \rf{rho-0-def}):
\be
p = - 2M_0 = {\rm const} \quad ,\quad
\rho = \rho_0 - p = 2M_0 + \Bigl(1+\frac{\P(C)}{\sqrt{-g}}\Bigr)\, 2Y 
\; ,
\lab{p-rho-def}
\ee

Because of the constant pressure ($p=-2M_0$) $\nabla^\n T_{\m\n}=0$ implies
{\em both} hidden Noether symmetry current $J^\m = \rho_0 u^\m$ conservation, 
as well as {\em geodesic fluid motion}:
\be
\nabla_\m \bigl(\rho_0 u^\m\bigr) = 0 \quad ,\quad 
u_\n \nabla^\n u_\m = 0 \; .
\lab{dust-geo}
\ee

Therefore, $T_{\m\n} = - 2M_0 g_{\m\n} + \rho_0 u_\m u_\n$ 
represents an exact sum of two contributions of the two dark species:
\br
p = p_{\rm DE} + p_{\rm DM} \quad,\quad \rho = \rho_{\rm DE} + \rho_{\rm DM}
\lab{DE+DM-1} \\
p_{\rm DE} = -2M_0\;\; ,\;\; \rho_{\rm DE} = 2M_0 \quad ; \quad
p_{\rm DM} = 0\;\; ,\;\; \rho_{\rm DM} = \rho_0 \; ,
\lab{DE+DM-2}
\er
\textsl{i.e.}, the dark matter component is a dust fluid flowing along
geodesics. This is explicit unification of dark energy and dark matter
originating from the dynamics of a single scalar field - the ``darkon'' $u$.

\section{Quintessential Inflation via Two Non-Riemannian Volume-Forms}
\label{quintess}

Let us now consider, following \ct{emergent}, a modified-measure gravity-matter theory
constructed in terms of two different non-Riemannian volume-forms 
(using again units where $G_{\rm Newton} = 1/16\pi$):
\be
S = \int d^4 x\,\P (A) \Bigl\lb R + L_1 (\vp,X)\Bigr\rb + 
\int d^4 x\,\P (B) \Bigl\lb L_2 (\vp,X) + 
\frac{\P (H)}{\sqrt{-g}}\Bigr\rb \; .
\lab{TMMT}
\ee
Here the following notations are used:
\begin{itemize}
\item
$\P(A)$ and $\P(B)$ are two independent non-Riemannian volume-forms:
\be
\P (A) = \frac{1}{3!}\vareps^{\m\n\k\l} \pa_\m A_{\n\k\l} \quad ,\quad
\P (B) = \frac{1}{3!}\vareps^{\m\n\k\l} \pa_\m B_{\n\k\l} \; ,
\lab{Phi-1-2}
\ee
\item
$\P (H) = \frac{1}{3!}\vareps^{\m\n\k\l} \pa_\m H_{\n\k\l}$  
is the dual field-strength of an additional auxiliary tensor gauge field $H_{\n\k\l}$
crucial for the consistency of \rf{TMMT}.
\item
We are using Palatini formalism: $R=g^{\m\n} R_{\m\n}(\G)$, where $g_{\m\n}$
and the affine connection $\G^\l_{\m\n}$  are {\em apriori} independent.
\item
$L_{1,2} (\vp,X)$ denote two different Lagrangians of a single scalar matter
field $\vp$ - the ``inflaton'', of the form:
\br
L_1 (\vp,X) = X - V_1 (\vp) \;\;, \;
X \equiv - \h g^{\m\n} \pa_\m \vp \pa_\n \vp \; ,\; V_1 (\vp) = f_1 \exp \{-\a\vp\} \; ,
\nonu
\phantom{aaaa}
\lab{L-1}
\er
\be
L_2 (\vp,X) = b e^{-\a\vp} X + U(\vp) 
\quad ,\quad U(\vp) = f_2 \exp \{-2\a\vp\} \; ,
\lab{L-2}
\ee
where $\a, f_1, f_2$ are dimensionful positive parameters, whereas $b$ is a
dimensionless one.
\item
The form of the action \rf{TMMT} is fixed by the requirement of invariance
under global Weyl-scale transformations:
\br
g_{\m\n} \to \l g_{\m\n} \;,\; \G^\m_{\n\l} \to \G^\m_{\n\l} \; ,\; 
\vp \to \vp + \frac{1}{\a}\ln \l\; ,
\nonu \\
A_{\m\n\k} \to \l A_{\m\n\k} \; ,\; B_{\m\n\k} \to \l^2 B_{\m\n\k} \; ,\; 
H_{\m\n\k} \to H_{\m\n\k} \; .
\lab{scale-transf}
\er
\end{itemize}

Equations of motion w.r.t. affine connection $\G^\m_{\n\l}$ yield a solution for
the latter as a Levi-Civita connection:
\be
\G^\m_{\n\l} = \G^\m_{\n\l}({\bar g}) = 
\h {\bar g}^{\m\k}\(\pa_\n {\bar g}_{\l\k} + \pa_\l {\bar g}_{\n\k} 
- \pa_\k {\bar g}_{\n\l}\) \; ,
\lab{G-eq}
\ee
w.r.t. to the Weyl-rescaled metric ${\bar g}_{\m\n}$:
\be
{\bar g}_{\m\n} = \chi_1 g_{\m\n} \quad ,\quad
\chi_1 \equiv \frac{\P_1 (A)}{\sqrt{-g}} \; . 
\lab{bar-g}
\ee
The metric ${\bar g}_{\m\n}$ plays an important role as the ``Einstein
frame'' metric (see \rf{einstein-frame} below).

Variation of the action \rf{TMMT} w.r.t. auxiliary tensor gauge fields
$A_{\m\n\l}$, $B_{\m\n\l}$ and $H_{\m\n\l}$ yields the equations:
\be
\pa_\m \Bigl\lb R + L^{(1)} \Bigr\rb = 0 \;, \;
\pa_\m \Bigl\lb L^{(2)} + \frac{\P (H)}{\sqrt{-g}}\Bigr\rb = 0 
\;, \; \pa_\m \Bigl(\frac{\P_2 (B)}{\sqrt{-g}}\Bigr) = 0 \; ,
\lab{A-B-H-eqs}
\ee
whose solutions read:
\br
\frac{\P_2 (B)}{\sqrt{-g}} \equiv \chi_2 = {\rm const} \quad ,\quad
R + L^{(1)} = M_1 = {\rm const} \; ,
\nonu \\
L^{(2)} +\frac{\P (H)}{\sqrt{-g}} = - M_2  = {\rm const} \; .
\lab{integr-const}
\er
Here $M_1$ and $M_2$ are arbitrary dimensionful and $\chi_2$
arbitrary dimensionless integration constants.

The first integration constant $\chi_2$ in \rf{integr-const} preserves
global Weyl-scale invariance \rf{scale-transf}
whereas the appearance of the second and third integration constants $M_1,\, M_2$
signifies {\em dynamical spontaneous breakdown} of global Weyl-scale invariance 
under \rf{scale-transf} 
due to the scale non-invariant solutions (second and third ones) in \rf{integr-const}. 

It is very instructive to elucidate the physical meaning of the three arbitrary 
integration constants $M_1,\, M_2,\,\chi_2$ from the point of view of the
canonical Hamiltonian formalism (for details, we refer to \ct{grf-essay}): 
$M_1,\, M_2,\,\chi_2$ are identified as conserved Dirac-constrained
canonical momenta conjugated to (certain components of) the auxiliary
maximal rank antisymmetric tensor gauge fields $A_{\m\n\l}, B_{\m\n\l}, H _{\m\n\l}$
entering the original non-Riemannian volume-form action \rf{TMMT}.

Performing transition from the original metric $g_{\m\n}$ to ${\bar g}_{\m\n}$ 
we arrive at the {\em ``Einstein-frame''}, where the gravity equations of motion 
are written in the standard form of Einstein's equations:
\be
R_{\m\n}({\bar g}) - \h {\bar g}_{\m\n} R({\bar g}) = \h T^{\rm eff}_{\m\n} 
\lab{einstein-frame}
\ee
with an appropriate {\em effective} energy-momentum tensor given in terms
of an Einstein-frame scalar Lagrangian $L_{\rm eff}$. The latter turns out
to be of the non-canonical ``k-essence'' (kinetic quintessence) type \ct{k-essence}
(containing higher powers of the scalar kinetic term ${\bar X}$:
\be
L_{\rm eff} = A(\vp) {\bar X} + B(\vp) {\bar X}^2 - U_{\rm eff}(\vp) \;\;\; ,
\;\; {\bar X} \equiv - \h {\bar g}^{\m\n} \pa_\m \vp \pa_\n \vp \; ,
\lab{L-eff-final}
\ee
where (recall $V_1 = f_1 e^{-\a\vp}$ and $U = f_2 e^{-2\a\vp}$):
\br
A(\vp) \equiv 1 + \h b e^{-\a\vp}
\frac{V_1 (\vp) + M_1}{U(\vp) + M_2} \;\; ,\;\;\;
B(\vp) \equiv - \frac{\chi_2 b^2 e^{-2\a\vp}}{4\bigl( U(\vp) + M_2\bigr)} \; ,
\lab{A-B-def} \\
U_{\rm eff} (\vp) \equiv 
\frac{(V_1 (\vp) + M_1)^2}{4\chi_2 \bigl( U(\vp) + M_2\bigr)} \; .
\phantom{aaaaaaaa}
\lab{U-eff}
\er

As a most remarkable feature, the effective scalar potential $U_{\rm eff}
(\vp)$ \rf{U-eff} possesses two {\em infinitely large flat regions}:

\begin{itemize}
\item
{\em (-) flat region} -- for large negative values of $\vp$, describing the
``early'' (inflationary) Universe:
\be
U_{\rm eff}(\vp) \simeq U_{(-)} \equiv \frac{f_1^2}{4\chi_2\,f_2} \; ,
\lab{U-minus} 
\ee
\item
{\em (+) flat region} -- for large positive values of $\vp$, describing the
``late'' (nowadays) Universe:
\be
U_{\rm eff}(\vp) \simeq U_{(+)} \equiv \frac{M_1^2}{4\chi_2\,M_2} \; ,
\lab{U-plus}
\ee
\end{itemize}

\begin{figure}
\begin{center}
\includegraphics[width=9cm,keepaspectratio=true]{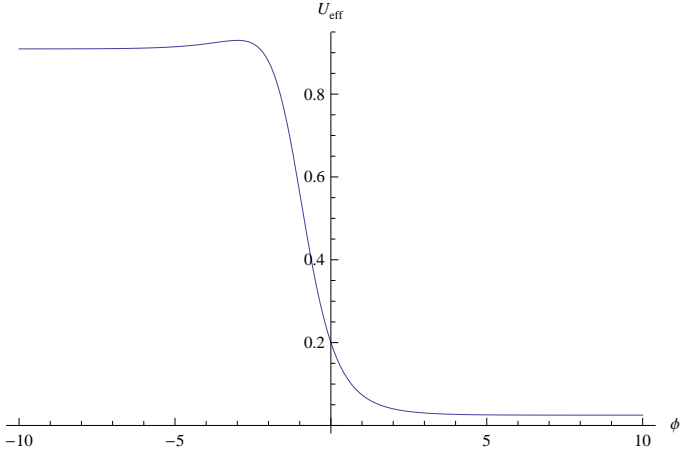}
\caption{Qualitative shape of the effective scalar potential $U_{\rm eff}$ \rf{U-eff}
as function of $\vp$.} 
\end{center}
\end{figure}

From the expression for $U_{\rm eff} (\vp)$ \rf{U-eff} and Fig.1 we
deduce that we have an {\em explicit realization of quintessential inflation scenario}
\ct{quintessence-orig}
-- continuously connecting an inflationary phase in the ``early'' Universe to a slowly 
accelerating expansion of ``present-day'' Universe \ct{accel-exp}
through the evolution of a single scalar field.

The flat regions \rf{U-minus} and \rf{U-plus} correspond indeed
to the evolution of the {\em ``early''} and the {\em ``late''} Universe, respectively, 
provided we choose the ratio of the coupling constants in the original scalar potentials 
versus the ratio of the scale-symmetry breaking integration constants to obey:
\be
\frac{f_1^2}{f_2} \gg \frac{M_1^2}{M_2} \, ,
\lab{early-vs-late}
\ee
which makes the {\em vacuum energy density of the early Universe $U_{(-)}$ much bigger
than that of the late Universe $U_{(+)}$} (cf. \rf{U-minus}, \rf{U-plus}).

If we choose the scales $M_1 \sim M^4_{EW}$ and $M_2 \sim M^4_{Pl}$, where 
$M_{EW},\, M_{Pl}$ are the electroweak and Plank scales, respectively, we are then 
naturally led to a very small vacuum energy density:
\be
U_{(+)}\sim M^8_{EW}/M^4_{Pl} \sim 10^{-120} M^4_{Pl} \; ,
\lab{U-plus-magnitude}
\ee
which is the right order of magnitude for the present epoch's vacuum energy
density as already realized in \ct{arkani-hamed}.

On the other hand, if we take the order of magnitude of the coupling
constants in the effective potential $f_1 \sim f_2 \sim (10^{-2} M_{Pl})^4$, 
then the order of magnitude of the vacuum energy density of the early Universe becomes:
\be
U_{(-)} \sim f_1^2/f_2 \sim 10^{-8} M_{Pl}^4 \; ,
\lab{U-minus-magnitude}
\ee
which conforms to the Planck Collaboration data \ct{Planck} 
implying the energy scale of inflation of order $10^{-2} M_{Pl}$.

\section{Quintessential Inflation and Unified Dark Energy and Dark Matter}
\label{quintess-DE-DM}

Now we will extend our results from the previous two sections by considering
a combination of the both models above \rf{TMMT} and \rf{TMT-0}
-- gravity coupled to both ``inflaton'' and ``darkon'' scalar fields within the
non-Riemannian volume-form formalism, as well as we will also add coupling to the
bosonic sector of the electro-weak model:
\br
S = \int d^4 x\,\P (A) \Bigl\lb g^{\m\n} R_{\m\n}(\G) + L_1 (\vp,X)
- g^{\m\n} \bigl(\nabla_\m \s_a)^{*}\nabla_\n \s_a - V_0 (\s)\Bigr\rb +
\nonu \\
\int d^4 x\,\P (B) \Bigl\lb L_2 (\vp,X)  
- \frac{1}{4g^2} F^2(\cA) - \frac{1}{4g^{\pr\,2}} F^2(\cB) +
\frac{\P (H)}{\sqrt{-g}}\Bigr\rb
\nonu \\
+ \int d^4 x \bigl(\sqrt{-g}+\P(C)\bigr) L(u,Y) \; .
\phantom{aaaaaaaaaa}
\lab{TMMT-1}
\er
Here we are using the same notations as in \rf{Phi-1-2}-\rf{L-2},  
\rf{mod-measure}-\rf{standard-L} and in addition:

\begin{itemize}
\item
$\s \equiv (\s_a)$ is a complex $SU(2)\times U(1)$ iso-doublet scalar field
with the isospinor index $a=+,0$ indicating the corresponding $U(1)$ charge.
The gauge-covariant derivative acting on $\s$ reads:
\be
\nabla_\m \s = 
\Bigl(\pa_\m - \frac{i}{2} \t_A \cA_\m^A - \frac{i}{2} \cB_\m \Bigr)\s \; ,
\lab{cov-der}
\ee
with $\h \t_A$ ($\t_A$ -- Pauli matrices, $A=1,2,3$) indicating the $SU(2)$ 
generators and $\cA_\m^A$ ($A=1,2,3$)
and $\cB_\m$ denoting the corresponding $SU(2)$ and $U(1)$ gauge fields.
\item
The ``bare'' $\s$-field potential is of the same form as the standard Higgs
potential:
\be
V_0 (\s) = \frac{\l}{4} \((\s_a)^{*}\s_a - \m^2\)^2 \; .
\lab{standard-higgs}
\ee
\item
The gauge field kinetic terms are (all indices $A,B,C = (1,2,3)$):
\br
F^2(\cA) \equiv F^A_{\m\n} (\cA) F^A_{\k\l} (\cA) g^{\m\k} g^{\n\l} \;\; ,\;\;
F^2(\cB) \equiv F_{\m\n} (\cB) F_{\k\l} (\cB) g^{\m\k} g^{\n\l} \; ,
\nonu \\
F^A_{\m\n} (\cA) = 
\pa_\m \cA^A_\n - \pa_\n \cA^A_\m + \eps^{ABC} \cA^B_\m \cA^C_\n \;\; ,\;\;
F_{\m\n} (\cB) = \pa_\m \cB_\n - \pa_\n \cB_\m \; .
\nonu \\
\er
\end{itemize}

Following the same steps as above, we derive from \rf{TMMT-1} the physical 
{\em Einstein-frame} theory w.r.t. Weyl-rescaled Einstein-frame metric 
${\bar g}_{\m\n}$ \rf{bar-g} and perform an additional ``darkon'' field redefinition
$u \to {\wti u}$: 
\be
\;\; \partder{{\wti u}}{u}=\bigl( V(u) - 2M_0\bigr)^{-\h} \quad ; \quad
Y \to {\wti Y} = - \h {\bar g}^{\m\n}\pa_\m {\wti u} \pa_\n {\wti u} \; .
\lab{darkon-redef}
\ee
The Einstein-frame action reads:
\be
S = \int d^4 x \sqrt{-{\bar g}} \Bigl\lb R({\bar g}) + 
L_{\rm eff}\bigl(\vp, {\bar X}, {\wti Y};\s,\cA,\cB\bigr)\Bigr\rb \; ,
\lab{einstein-frame-1}
\ee
where (recall ${\bar X} = - \h {\bar g}^{\m\n}\pa_\m \vp \pa_\n \vp$):
\br
L_{\rm eff}\bigl(\vp, {\bar X}, {\wti Y};\s,\cA,\cB\bigr) = {\bar X} 
- {\wti Y}\Bigl(V_1(\vp) + V_0 (\s) + M_1 - \chi_2 b e^{-\a\vp}{\bar X}\Bigr) 
\nonu \\
+ {\wti Y}^2 \Bigl\lb \chi_2 (U(\vp) + M_2) - 2 M_0\Bigr\rb + L\lb\s,\cA,\cB\rb \; ,
\phantom{aaaaaaaa}
\lab{L-eff-total} 
\er
with:
\be
L\lb\s,\cA,\cB\rb \equiv
- {\bar g}^{\m\n} \bigl(\nabla_\m \s_a)^{*}\nabla_\n \s_a
- \frac{\chi_2}{4g^2} {\bar F}^2(\cA) - \frac{\chi_2}{4g^{\pr\,2}} {\bar F}^2(\cB) \; .
\lab{L-sigma-gauge-def} 
\ee

Tha Lagrangian \rf{L-eff-total}  is again of a generalized ``k-essence'' form 
(non-linear w.r.t. both ``inflaton'' and ``darkon'' kinetic terms ${\bar X}$ and 
${\wti Y}$). $M_0$ and $M_1, M_2, \chi_2$ are the same integration constants as in
\rf{L-const} and \rf{integr-const}, respectively.

The action \rf{einstein-frame-1}-\rf{L-eff-total} possesses an obvious Noether
symmetry under the shift ${\wti u} \to {\wti u} + {\rm const}$ with
current conservation:
\be
\pa_\m \Bigl(\sqrt{-{\bar g}} {\bar g}^{\m\n}\pa_\n {\wti u} 
\partder{L_{\rm eff}}{\wti Y}\Bigr) = 0 \;\; ,
\lab{J-conserv-EF}
\ee
which is Einstein-frame counterpart of the original $g_{\m\n}$-frame
``dust'' dark matter density conservation \rf{J-conserv}.

For static (spacetime idependent) scalar field configurations (here the original 
``darkon'' field $u$ is static, whereas the transformed one ${\wti u}$ \rf{darkon-redef}
is {\em not} -- this is due to the dynamical Lagrangian ``darkon'' constraint 
\rf{L-const}) we have:
\be
{\wti Y}\bv_{\rm static} = \frac{V_1 (\vp) + V_0 (\s) + M_1}{
2 \chi_2 \bigl( U(\vp) + M_2\bigr) - 4 M_0} \; ,
\lab{Y-static}
\ee
which upon substitution into \rf{L-eff-total} yields the following total scalar field
effective potential (cf. Eq.\rf{U-eff}):
\be
U_{\rm eff}\bigl(\vp,\s\bigr) = 
\frac{\Bigl(V_1 (\vp) + V_0 (\s) + M_1\Bigr)^2}{4\bigl\lb \chi_2 (U(\vp) +
M_2) - 2 M_0\bigr\rb}
\lab{U-eff-total}
\ee
As for the purely ``inflaton'' potential \rf{U-eff}, the ``inflaton+Higgs''
potential \rf{U-eff-total} similarly possess two infinitely large regions: $(-)$ flat
region for large negative and $(+)$ flat region and large positive values of the
``inflaton'', respectively, as in Fig.1 (when $\s$ is fixed).

\begin{itemize}
\item 
In the $(+)$ flat region \rf{U-eff-total} reduces to (cf. \rf{U-plus}):
\be
U_{\rm eff}\bigl(\vp,\s\bigr) \simeq U_{(+)}(\s) =
\frac{\Bigl(\frac{\l}{4} \((\s_a)^{*}\s_a - \m^2\)^2 + M_1\Bigr)^2}{
4\bigl(\chi_2 M_2 - 2 M_0\bigr)} \; ,
\lab{U-plus-higgs}
\ee
which obviously yields as a lowest lying vacuum the Higgs one:
\be
|\s| = \m  \; ,
\lab{higgs-vac}
\ee
\textsl{i.e.}, in the ``late'' (post-inflationary) Universe we have the
standard spontaneous breakdown of $SU(2)\times U(1)$ ~gauge symmetry.
Moreover, at the Higgs vacuum \rf{higgs-vac} we obtain from \rf{U-plus-higgs}
a dynamically generated cosmological constant $\L_{(+)}$ of the ``late'' Universe:
\be
U_{(+)}(\m) \equiv 2\L_{(+)} = \frac{M_1^2}{4\bigl(\chi_2 M_2 - 2 M_0\bigr)} \; .
\lab{CC-eff-plus}
\ee
\item
In the $(-)$ flat region \rf{U-eff-total} reduces to the same expression
as in \rf{U-minus}, which is $\s$-field idependent. Thus, the Higgs-like
iso-doublet scalar field $\s_a$ remains {\em massless} in the ``early''
(inflationary) Universe and accordingly there is {\em no} electro-weak
spontaneous symmetry breaking there.
\end{itemize}

To study cosmological implications of \rf{TMMT-1} we perform a
Friedmann-Lemaitre-Robertson-Walker (FLRW) reduction to the class of FLRW metrics:
\be
ds^2 = {\bar g}_{\m\n} dx^\m dx^\n = - N^2(t) dt^2 + a^2(t) d{\vec x}.d{\vec x}
\lab{FLRW}
\ee
and take the ``inflaton'' and ``darkon'' to be time-dependent only:
\be
{\bar X} = \h \vpdot^2 \quad,\quad {\wti Y} = \h v^2 \;\;, \;\;
v \equiv \frac{d{\wti u}}{dt} \; .
\lab{X-Y-FLRW}
\ee
Upon variation w.r.t. ``lapse'' $N(t)$ we take the usual gauge $N(t)=1$.

Now, the FLRW reduction of the ``darkon'' ${\wti u}$-eqs. of motion \rf{J-conserv-EF}
yields a {\em cubic algebraic} eq. for its velocity $v$:
\be
\Bigl\lb \chi_2 (U(\vp) + M_2) - 2 M_0\Bigr\rb v^3 - v \Bigl(V_1(\vp) + V_0 (\s) + M_1
- \chi_2 b e^{-\a\vp}\h\vpdot^2\Bigr) - \frac{c_0}{a^3} = 0  \; ,
\lab{w-cubic}
\ee
where $c_0$ is an integration constant -- the conserved Noether charge of
\rf{J-conserv-EF} (``dust'' dark matter particle number).

The equations of motion w.r.t. $N(t)$ and $a(t)$ (1st and 2nd Friedmann eqs.) read:
\be
\frac{\adot^2}{a^2} = \frac{1}{6}\,\rho \quad ,\quad
\frac{\addot}{a}= -\frac{1}{12}\bigl(\rho + 3 p\bigr) \; ,
\lab{friedman-eqs}
\ee
where the energy density $\rho$ and pressure $p$ are given by:
\br
\rho = \h\vpdot^2 \Bigl(1+\frac{3}{4}\chi_2 b e^{-\a\vp} v^2\Bigr)
+\frac{v^2}{4}\bigl(V_1(\vp) + V_0 (\s) + M_1\bigr) + \frac{3}{4}\,\frac{c_0}{a^3}\, v \; ,
\lab{rho-def} \\
p = \h\vpdot^2 \Bigl(1+\frac{1}{4}\chi_2 b e^{-\a\vp} v^2\Bigr)
- \frac{v^2}{4}\bigl(V_1(\vp) + V_0 (\s) + M_1\bigr) + \frac{1}{4}\,\frac{c_0}{a^3}\, v \; .
\lab{p-def}
\er
Finally, the equation of motion w.r.t. ``inflaton'' $\vp$ reads:
\br
0=\frac{d}{dt} \Bigl\lb a^3 \vpdot \Bigl( 1 +
\frac{\chi_2}{2} b e^{-\a\vp}v^2\Bigr)\Bigr\rb + 
\a\frac{\chi_2\, U(\vp)\, c_0\, v}{\chi_2\bigl(U(\vp)+M_2\bigr) - 2 M_0} + 
\nonu \\
\a a^3 v^2 \Bigl\{ \frac{\vpdot^2}{4} \chi_2 b e^{-\a\vp} 
- \h \bigl( V_1(\vp) + V_1(\s)\bigr) + 
\nonu \\
\frac{\chi_2 U(\vp)\llb V_1(\vp) + V_0 (\s) + M_1 - 
\chi_2 b e^{-\a\vp}\vpdot^2\!\!/2\rrb}{2 \llb\chi_2\bigl( U(\vp)+M_2\bigr) 
- 2 M_0\rrb} \Bigr\} 
\; .
\lab{vp-eq}
\er

First, let us consider the $(+)$ flat region \rf{U-plus} of the inflaton potential
\rf{U-eff-total} (right flat region on Fig.1)
for large positive values of $\vp$ corresponding to the ``late'' (nowadays) Universe.
In this case we have from \rf{w-cubic}, \rf{rho-def} and \rf{p-def} (taking
into account \rf{standard-higgs} and \rf{higgs-vac}):
\br
v = \Bigl\lb\frac{M_1}{\chi_2 M_2 - 2 M_0}\Bigr\rb^{\h} +
\frac{1}{2M_1}\, \frac{c_0}{a^3} + {\rm O} \bigl(\frac{c^2_0}{a^6}\bigr)
\; ,
\lab{w-plus} \\
\rho = \frac{M_1^2}{4(\chi_2 M_2 - 2 M_0)} + \frac{c_0}{a^3}\,
\Bigl\lb\frac{M_1}{\chi_2 M_2 - 2 M_0}\Bigr\rb^{\h} 
+ {\rm O} \bigl(\frac{c^2_0}{a^6}\bigr) \; ,
\lab{rho-plus} \\
p = - \frac{M_1^2}{4(\chi_2 M_2 - 2 M_0)} + 
{\rm O} \bigl(\frac{c^2_0}{a^6}\bigr) \; .
\lab{p-plus}
\er
Substituting \rf{rho-plus} into the first Friedmann Eq.\rf{friedman-eqs} we
obtain (the solution for $a(t)$ below first appeared in \ct{turner-etal}):
\be
a(t) \simeq \Bigl(\frac{{\wti C}_0}{2\L_{(+)}}\Bigr)^{1/3} 
\sinh^{2/3}\Bigl(\sqrt{\frac{3}{4}\L_{(+)}}\, t\Bigr) \;\; ,\;\;
\vpdot \simeq {\rm const}\, \sinh^{-2}\Bigl(\sqrt{\frac{3}{4}\L_{(+)}}\, t\Bigr) \; ,
\lab{a-plus}
\ee
with $\L_{(+)}$ as in \rf{CC-eff-plus} and
${\wti C}_0 \equiv c_0 \sqrt{M_1 (\chi_2 M_2 - 2 M_0)^{-1}}$.

Relations \rf{rho-plus} and \rf{p-plus} straightforwardly show that in the ``late''
(nowadays) Universe we have explicit unification of dark energy (given by
the dynamically generated cosmological constant \rf{CC-eff-plus} -- first
terms on the r.h.s. of \rf{rho-plus} and \rf{p-plus}), and dark matter given
as a ``dust'' fluid contribution -- second term on the r.h.s. of \rf{rho-plus}.

Next consider the $(-)$ flat region \rf{U-minus} of the inflaton potential
\rf{U-eff-total} (left flat region on Fig.1) for large negative values of $\vp$
corresponding to the ``early'' (``inflationary'') Universe. We will consider
the ``slow-roll'' inflationary epoch \ct{slow-roll}
(\textsl{i.e.}, $\vpddot, \vpdot^2, \vpdot^3, \ldots$ -- ignored) :
\br
v = e^{\h\a\vp}\Bigl\lb v_1 + \frac{1}{2f_1}\,\frac{c_0}{a^3}\, e^{\h\a\vp} 
+ {\rm O}\bigl(e^{\a\vp}\bigr)\Bigr\rb \quad ,\;\; 
v_1 \equiv - \Bigl(\frac{f_1}{\chi_2 f_2}\Bigr)^{\h} \; ,
\lab{w-minus} \\
\rho = U_{(-)} - e^{\h\a\vp} |v_1|\,\frac{c_0}{a^3} + \frac{1}{4} {\wti M} v_1^2 e^{\a\vp}
+ {\rm O}\Bigl(e^{3\a\vp/2}, \frac{c_0^2}{a^6}\Bigr) \; ,
\lab{rho-minus} \\
U_{(-)} \equiv \frac{f_1^2}{4\chi_2 f_2} \;\; ,\;\;
{\wti M} \equiv M_1 + V_0 (\s=0) = M_1 + \frac{\l}{4}\m^4 \; .
\lab{M-wti}
\er
Friedmann \rf{friedman-eqs} and inflaton \rf{vp-eq} equations can be solved
analytically in the ``slow-roll'' approximation for the special relation among
parameters $1+ \frac{b f_1}{2f_2}= \frac{2}{3}\a^2$. 
In the latter case we have:
\be
\vpdot - \frac{|v_1|}{2\a H_0} \Bigl\lb \frac{c_0}{c_1^3}\, e^{-3H_0 t}
  -\frac{1}{4} {\wti M} |v_1|\Bigr\rb\, e^{\a\vp} = 0 \quad , \quad 
H_0 \equiv \sqrt{\frac{1}{6}U_{(-)}} \; ,
\lab{vpdot-eq}
\ee
where $c_1$ is another integration constant. For the inflaton field and
Friedmann scale factor we obtain:
\br
e^{-\a\vp (t)} = c_2 + \frac{|v_1|}{2\a H_0} 
\Bigl(\frac{c_0}{3c_1^3 H_0}\, e^{-3H_0 t} + \frac{1}{4} {\wti M} |v_1|\, t\Bigr) \; ,
\lab{vp-minus} \\
a(t) = c_1 e^{H_0 t}\, e^{-\frac{1}{6}\a\vp (t)} \; ,
\lab{a-minus}
\er
where $c_2$ is a third integration constant.

Eqs.\rf{vpdot-eq}-\rf{a-minus} display the effect of the presence of
``dusty'' dark matter ($c_0 \neq 0$) on the ``slow-roll'' inflationary
evolution (here we must have $\vpdot \geq 0$):

\begin{itemize}
\item
$\vpdot (t) >0$ for 
$t<t_{*} \equiv \frac{1}{3H_0}\,\ln\bigl(4c_0 ({\wti M} |v_1| c_1^3)^{-1}\bigr)$,
where $\vpdot (t_{*})=0$, \textsl{i.e.}, $\vp (t)$ rolls forward untill
$t=t_{*}$.
\item
According to \rf{a-minus} the prefactor $e^{-\frac{1}{6}\a\vp (t)}$ 
of the inflationary time exponential $e^{H_0 t}$ drops down with $t \leq t_{*}$.
\end{itemize}

For $t > t_{*}$ the evolution described by the inflaton solution
\rf{vp-minus} {\em cannot anymore} be valid, since according to \rf{vpdot-eq}
the inflaton velocity is negative for $t > t_{*}$, \textsl{i.e.}, for $t > t_{*}$
the inflaton would start rolling backwards. This non-validity of \rf{vp-minus}
is due to the fact that for
$t \sim t_{*}$ the inflaton value $\vp(t)$ exits the $(-)$ flat region of
the inflaton effective potential \rf{U-eff-total} (left flat region on Fig.1).
The latter sets the following constraint on the integration constant $c_2$ in
\rf{vp-minus} for the latter to be valid:
\be
e^{-\a \vp(t_{*})} \equiv c_2 + \frac{{\wti M}}{f_1}\Bigl\lb 1 + 
\ln\Bigl(\frac{4c_0}{{\wti M} |v_1| c_1^3}\Bigr)\Bigr\rb <
\frac{f_1 (\chi_2 M_2 - 2 M_0)}{f_2 \chi_2 {\wti M}} 
\equiv e^{-\a \vp_{\rm max}} \; ,
\lab{c2-constr}
\ee
where $\vp_{\rm max}$ is the location of the maximum of the inflaton
potential \rf{U-eff-total} -- the small ``bump'' on the left half of Fig.1,
which is just outside the $(-)$ flat region.

Let us note that the relative height $\Delta U_{(-)}$ of the above mentioned 
``bump'' of the inflaton potential \rf{U-eff-total} w.r.t. the height of the $(-)$ 
flat region \rf{U-minus}:
\be
\Delta U_{(-)} \equiv U_{\rm eff} (\vp_{\rm max},\m) - \frac{f_1^2}{4\chi_2 f_2}
= \frac{{\wti M}^2}{4\bigl(\chi_2 M_2 - 2 M_0\bigr)}
\lab{delta-bump}
\ee
(${\wti M}$ as in \rf{M-wti}) is of the same order of magnitude as the small effective
cosmological constant \rf{CC-eff-plus} in the $(+)$ flat region (``late'' Universe).

\section{Conclusions}
\label{conclude}

The non-Riemannian volume-form formalism (\textsl{i.e.}, employing alternative 
non-Riemannian reparametrization covariant integration measure densities on the 
spacetime manifold) has substantial impact in any general-coordinate or
reparametrization invariant field theories.

\begin{itemize}
\item
The non-Riemannian volume-form formalism in gravity/matter theories
naturally provides a self-consistent unified description of dark energy 
as dynamically generated cosmological constant and dark matter as a ``dust''
fluid flowing along geodesics realized through the dynamics of a single ``darkon'' 
scalar field. This unification becomes manifest within the ``late'' (dark
energy dominated) epoch of the Universe's evolution.
\item
Employing two different non-Riemannian volume-forms leads to the construction of a
new class of ``quintessential'' gravity-matter models, producing an effective scalar 
``inflaton'' potential with two infinitely large flat regions. This allows for a 
unified description of both early Universe inflation as well as of present dark 
energy dominated epoch. 
\item
The above non-conventional ``quintessential'' gravity-matter models can be
extended to include both the ``darkon'' as well as the fields comprising the 
bosonic sector of the electroweak theory, in
particular -- a Higgs-like scalar $\s$, whereby producing {\em dynamically} in the
post-inflationary epoch an effective potential for $\s$ of the  canonical electroweak 
symmetry breaking Higgs form, while keeping the electroweak gauge symmetry
intact in the early inflationary Universe. 
\end{itemize}

Let us also note that application of the non-Riemannian volume-form
formalism in the context of minimal $N=1$ supergravity \ct{susyssb} naturally generates a 
{\em dynamical cosmological constant} as an arbitrary dimensionful integration 
constant, which triggers {\em spontaneous supersymmetry breaking} and mass generation for 
the gravitino -- a new mechanism for the {\em supersymmetric Brout-Englert-Higgs effect}.

\section*{Acknowledgements}
We gratefully acknowledge support of our collaboration through the academic exchange 
agreement between the Ben-Gurion University and the Bulgarian Academy of Sciences.
S.P. and E.N. have received partial support from European COST actions
MP-1210 and MP-1405, respectively, as well from Bulgarian National Science
Fund Grant DFNI-T02/6. E.G. received partial support from COST Action CA-15117.


\end{document}